# Nothing's plenty: The significance of null results in physics education research

Luke D. Conlin[1], Eric Kuo[2], and Nicole R. Hallinen[3]
[1]*Department of Chemistry and Physics, Salem State University, Salem, MA 01970, USA*
[2]*Learning Research & Development Center, University of Pittsburgh, Pittsburgh, PA 15260, USA*
[3]*College of Education, Temple University, Philadelphia, PA 19122, USA*

A central aim of physics education research is to understand the processes of learning and use that understanding to inform instruction. To this end, researchers often conduct studies to measure the effect of classroom interventions on student learning outcomes. Many of these intervention studies have provided an empirical foundation of reformed teaching techniques, such as active engagement. However, many times there is not sufficient evidence to conclude that the intervention had the intended effect, and these null results often end up in the file drawer. In this paper, we argue that null results can contribute significantly to physics education research, even if the results are not statistically significant. First, we review social sciences and biomedical research that has found widespread publication bias against null results, exploring why it occurs and how it can hurt the field. We then present three cases from physics education research to highlight how studies that yield null results can contribute to our understanding of teaching and learning. Finally, we distill from these studies some general principles for learning from null results, proposing that we should evaluate them not on whether they reject the null hypothesis, but according to their potential for generating new understanding.

## I.  INTRODUCTION

What research findings are significant enough to be published? One answer to this question is that the criterion for publication should be whether a study advances our understanding of educational phenomena in teaching and learning. Yet, at times, research studies that meet this criterion fail to get published, because they present no statistically significant results. The issue is that there is a publication bias against null results.

This problem is not unique to educational research. Recent surveys have found evidence of widespread publication bias in social science [1,2] and biomedical literature [3]. These surveys have found that publication bias often results from an overemphasis by authors and journals alike on the p-value threshold of statistical significance (e.g., p<.05). No matter the discipline, withholding publication because of a p-value does damage to scientific progress. It keeps important insights from being disseminated, and further conflates statistical significance with real world significance [4].

In this survey article, we draw on literature from within PER and beyond to distill several lessons that can be learned from null results. We begin by describing an insightful null result in physics, the Michelson-Morley experiment. We then describe more carefully what we mean by a null result, and describe the evidence for a widespread publication bias against null results in the social sciences and medical research fields. Then turning our attention to physics education research, we then present examples of null results from our own research, highlighting the ways in which such null results can enhance our understanding of the educational phenomena underlying teaching and learning. Finally, we propose an initial set of guidelines for learning from null results, and discuss their implications for both research and instruction.

## II.  WHAT ARE NULL RESULTS AND WHAT DO WE DO WITH THEM?
### A.  A famous null result from physics: The Michelson-Morley experiment

One of the most influential physics experiments was a null result. In 1887, Michelson and Morley [5] published their empirical attempt to measure the relative motion between the Earth and the luminiferous ether, the theoretical medium through which light was said to travel.  At the time, the ether was thought to be stationary in space as the Earth moved through it, meaning that the speed of light measured by an observer on the Earth would be transformed due to that relative motion.  Light traveling parallel to the velocity of the Earth would appear slower, and light traveling anti-parallel to the Earth would appear faster.  Michelson and Morley used an interferometer to measure the phase shift between two beams of light traveling in perpendicular directions.   Although they made these measurements at different angles and different times of day, they did not find the predicted effect.  They concluded that their measurements indicated the phase shift was probably less than 1/40th of the predicted value based on the properties of the experimental apparatus and the velocity of the surface of the Earth.  Although counter to their predictions, Michelson and Morley pointed out that these results, if true, contradicted the theoretical models of light the time.  To explain this null result, they proposed one potential explanation for why their experiment failed: the motion of the solar system, which was not taken into account in their design, could have been exactly right at the time of the experiment to cancel out the motion of the Earth relative to the solar system.   Although they argued that this explanation seemed unlikely, they proposed to repeat the experiment every three months, so that this explanation can be ruled out.  Although these experiments failed to find the hypothesized effect, the null findings played a key role in Einstein's rejection of both the ether and a radical reformulation of relative motion in his Special Theory of Relativity nearly 20 years later.

   This exceptional case from physics illustrates how null results can play an important role in scientific innovation. Here, the theory made a strong prediction, and a well-designed experiment that should have detected a result failed to do so.  Instead of considering their experiment a failure and discarding the result, the scientists published the null result, making it public so that the result could become part of discussions in the field.  Michelson and Morley carefully tried to understand the reasons for the null result, proposing a new hypothesis that was carried out in subsequent experiments.  The continued failure, by Michelson, Morley, and others, to find the hypothesized effect challenged existing models and pushed the field to develop new ideas, which in this case contributed to the start of modern physics.

   Analogically, we use this example of a productive null result in physics to motivate the argument for productive null results in education research.  In the same way as Michelson and Morley, we believe that educational researchers can contribute to a growing body of knowledge - in this case, knowledge about teaching and learning - with surprising null results derived through well-designed research studies. Null results can challenge existing educational models and spur theoretical and methodological innovation.  Admittedly, there are many differences between modern research in physical sciences and the social sciences, including different senses of experimental control, different contributors to measurement error and uncertainty, and different levels of theoretical precision and understanding.  Still, as we will illustrate, the case for valuing productive null results in research generalizes across disciplines.

### B. What is a null result?

Quantitative hypotheses in education research often predict an association between two or more variables. Paired with every research hypothesis is a null hypothesis, which predicts no association between these variables.  For example, in comparing active learning approaches to traditional, lecture-style instruction in physics teaching, one might hypothesize an association between instructional approach and concept inventory score (e.g., active learning instruction will produce higher concept inventory scores).  Here, the null hypothesis would predict no difference in concept inventory scores for the two approaches. Similarly, within a physics course, a researcher might hypothesize that students' surveyed expectations toward learning physics (measured, say, through their MPEX score [6]) would positively correlate with their course grade.  In this case, the null hypothesis would predict no correlation between MPEX score and course grade.

Although seemingly redundant, the importance of null hypotheses arises due to the approach to statistical testing in educational research. Commonly, educational hypotheses predict associations between variables without predicting the magnitude of those associations. In contrast, hypotheses in experimental physics often make exact predictions of the magnitude of variable associations through theoretical calculations or simulations. Because of this inability to make exact predictions, educational hypotheses are rarely tested directly. Rather, analysis commonly engages in *null hypothesis significance testing*, to determine whether the measured values are statistically different from the null hypothesis of zero association between variables. If the associations found in the data are unlikely to have come from the null hypothesis (commonly, "unlikely" is defined as a less than 5% chance that an association of at least that magnitude could have been sampled by chance), then the null hypothesis is rejected, and we claim that the data support the original research hypothesis. A null result is when the null hypothesis cannot be rejected, and it remains statistically possible that there is no association between variables.

The impact of null results depends on the state of knowledge in the field. Null results are not surprising in situations where we did not expect associations between quantities to exist. However, if a null result contradicts commonly accepted theoretical models or intuitive beliefs by failing to find something that was expected, as was the case for the Michelson-Morley experiment, it may present an important challenge to existing theories or experimental methods. Similarly, a null result could contradict previous experimental results, suggesting that there is more to be understood about the designs of the experiments or about the theory connecting them. These surprising null results are the focus of this paper.

To summarize our argument, it is the nature of science to seek new understanding and making surprising results public can facilitate this process in research communities. A recent example from physics is the proposed detection of faster-than-light neutrinos by the OPERA experiment [7]. This finding contradicted a coherent and predictive theory of particle physics that is backed by a significant body of empirical results. Still, researchers published their surprising result, which initiated efforts to replicate and understand their findings. Several other research teams failed to replicate the findings, and eventually the cause of the original result was determined - an error in the experimental apparatus. Although this case was not one where a surprising result led to theoretical innovation, it illustrates the process of how scientific research fields can engage with such results. Rather than discarding or ignoring a result that did not fit with the existing state of knowledge, here scientists worked to replicate and investigate the surprising result until its mismatch with existing knowledge could be resolved.

This example also demonstrates the importance of replication. Because surprising findings may result from experimental error, replication can increase confidence that the surprising result represents an actual phenomenon to be understood. Replication also helps support the generalizability and reliability of the finding. Many educational effects could plausibly depend on the student populations, the local course and school contexts, and implementation details. Replication of educational effects across contexts bolsters claims for the stability and generality of the claims. For example, Freeman et al. [9] performed a meta-analysis of 225 studies of active learning instruction. They found that the benefits of active learning instruction over traditional methods did not differ by STEM discipline, majors versus non-majors courses, or introductory versus upper-division courses, and the effects were held for course exams, standardized concept inventories, and a range of class sizes. These replications show that active learning is a powerful educational prescription which is effective in a variety of course contexts and for a variety of outcome measures. In contrast, null results that indicate failures to replicate an effect are important for defining the bounds around which these effects are generalizable, reliable, and practically useful.

### C. The "file drawer" problem and its consequences

Despite the arguments that null results have a key role to play in the scientific process and in sharpening our understanding, null results are often not shared publicly. One major reason is that there is a publication bias in favor of statistically significant results. Null results are often viewed by journal editors or reviewers as "non-findings," and this perspective can be a significant barrier to publication.

Rosenthal [8] discussed this "file drawer problem," that null results are often filed away privately in researchers' file drawers, stating the most extreme possibility for this problem: The research literature is comprised of a biased sample of studies; the significant findings in the literature are the 5% of studies that show differences by chance, and the remaining 95% of the studies that cannot reject the null hypothesis remain unpublished. This publication bias is not just theoretical. A robust finding across surveys of published research is that almost every study reports statistically significant results. Sterling, Rosenbaum, and Weinkam [10] found that 95.6% of social science journal articles rejected the null hypothesis, as did 85.4% of medical journal articles. A more recent textual review [3] of millions of biomedical journal abstracts and full-text articles found and even higher proportion of studies rejecting the null hypothesis, with 96% of p-values reporting statistically significant results. It is unclear whether this is because the true proportion of significant findings rose, or whether the sample of studies that have been published became more biased against null results.

It is difficult to determine the full extent of the file drawer problem, since a full accounting would need to determine the proportion of unpublished papers that were null results. Recent approaches, such as the Time-sharing Experiments in Social Science (TESS), have increased transparency by subjecting research methods to peer review before data is collected, which allow for a more careful accounting of the fate of studies, whether published or not. Franco, Malhotra, and Simonovits [1] used TESS to analyze a set of 221 social sciences studies in whose fates were known to be either published, unpublished, or not written up at all. They found that studies characterized as having "strong" results were 40 times more likely to be published, and 60 times more likely to be written up. One question this brings up: What counts as a strong result, and to whom? Often, to both journals and authors alike, a key indicator of strength is a p-value of less than 0.05. This points to a root cause of publication bias against null results -- an overemphasis on hypothesis testing using a p-value cutoff [2].

The use of a p-value cutoff has been losing favor as the primary criterion of significant research findings within social science and biomedical science. The use of a p-value to judge the statistical significance of results (and the use of p=.05 specifically) was popularized in statistics by Fisher [11]. However, Fisher himself acknowledged the arbitrary nature of such a cutoff and later even disavowed himself of the practice of hypothesis testing with a set p-value as a general decision cutoff [12]:

> *"...the calculation is absurdly academic, for in fact no scientific worker has a fixed level of significance at which from year to year, and in all circumstances, [s]he rejects hypotheses; [s]he rather gives [her] mind to each particular case in the light of [her] evidence and [her] ideas."*

Following many critiques of the practice, particularly since the 1990s [4,13], the American Psychological Association has revised their research guidelines to de-emphasize p-values as the sole criterion of statistical effects, increasing emphasis on effect sizes and measures of uncertainty. The American Statistical Association has more recently put out a clarifying statement on the use of p-value cutoffs for hypothesis testing [14], arguing that "no single index should substitute for scientific reasoning." Despite reasons not to use p-value cutoffs as the sole measure of research quality, the reporting of significant p-values still holds sway with journals and authors alike. In their review of millions of biomedical journal abstracts and full-text articles, Ioannides and colleagues [3] found that the percentage of articles reporting of p-values has increased dramatically over time, doubling from 1990 to 2015.

A major problem created by the file drawer problem is an inability to estimate the replicability of published effects. Because the culture of science values innovation over replication, high-fidelity replication attempts are devalued, no matter the replication outcome. If the finding is not replicated, it may be devalued, because it does not present a statistically significant finding. On the other hand, if the finding is replicated, it may not be accepted for publication, because the finding is not novel. This constitutes a "reverse file drawer problem", where effects may be more reliable than depicted in the literature, because unpublished, successful replications exist. Yet, the attitudes toward these replications belie their importance as they can address a very real possibility that published findings may not be reliably replicated. Failure to replicate is a well-known problem in the social sciences as well as biomedical research [15]. Recently, the Open Science Collaborative performed high-quality replication attempts for 100 psychology studies. Only 39% of these attempts were rated to have replicated the original findings.

Meta-analyses provide a useful way of combining findings from a set of studies to evaluate the reliability and average magnitude of an effect, but, again, the publication bias in favor of significant results likely leads to overestimates of reliability and effect size. Researchers conducting meta-analyses can deal with these biases in several ways. One is to seek out effects in unpublished papers (such as dissertations) or to even contact researchers in the field to obtain results from unreported studies. Another is to calculate how a number of unpublished null results with effect size = 0 would affect the results. For example, in a meta-analysis of 158 studies comparing conceptual inventory scores after traditional instructional methods and after active learning in STEM classrooms [9], the size of the effect of active learning on exams and concept inventories was 0.47 SD. To investigate the potential effect of publication bias, they calculated the number of studies with an effect size of zero needed to reduce the overall effect size to 0.20 SD, finding that 114 such unpublished null results would be needed. Although this provides a useful estimate of how unpublished studies would affect the results, there are unfortunately no clear guidelines to determine the plausible number of null result studies that are unpublished.

The file drawer problem may face additional challenges in PER, which in addition to being a research field is also a practical effort to promote the use of its designed teaching methods by instructors and departments. Physics Education Researchers may be reluctant to publicize failures to replicate the efficacy of its instructional innovations for fear that resistant faculty may use such null results to justify their refusal to change their instructional approach. For this reason, PER may face additional reasons why it is undesirable to share null results.

### III. HOW PUBLISHING NULL RESULTS CAN CONTRIBUTE TO KNOWLEDGE IN EDUCATION RESEARCH

There are two alternatives to filing null results away, each of which can lead to innovations in the research field. One possibility is that the null results can be used to inspire innovation locally for the researchers who find them, even without publishing right away. That is, null results can lead the researchers who find them to reconsider and revise the research study so that significant effects between variables can be shown in subsequent trials. While this is locally a positive outcome for innovation from null results, a risk is that this innovation may not be adequately communicated in the literature. The original null results may not be published, and so the critical revisions are underemphasized, treated as experimental tweaking rather than critical elements of the experimental design and the empirical findings.

The other alternative to filing null results away is to publish them. In many cases, educational research studies may require significant time, effort, and access to classroom populations, so attempts to revise and repeat the study may not be possible. When local innovation is not possible, publishing the null results can invite global innovation by allowing the research community to participate in revision of theoretical ideas or methodological designs. Communicating null results makes public avenues of work that are unproductive, reducing duplication of research effort. Additionally, other researchers may have different perspectives on null results and have ideas for revising the experimental design or theoretical models. In this way, null results can be generative sources of new research directions.

Again, the argument is that research studies should not be judged solely on whether the quantitative results produce significant p-values. The significance of research studies should be judged according to whether or not they contribute to the state of understanding in the field. In particular, null results, which are often judged as not novel or non-findings, can play an important role by challenging existing ideas, advancing collective understanding of the mechanisms, boundaries, and applications of our theories.

In the following section, we present null results from our own work and describe how we think these null results can contribute to understandings of educational phenomena. We highlight these studies not as exemplars, so much as case studies of how researchers in PER can learn from null results.

### A. Null results can serve as existence proofs: the case of the general principle strategy

Null results can "surface" ideas that are not yet widely considered. Research can help map out the research space by providing an existence proof that a novel alternative approach may be viable, even when the outcomes produced are not measurably different than current approaches. This was the case in a study by Conlin, Hallinen, and Schwartz [16], which compared how middle school students learned and used a new strategy for conducting scientific inquiry.

Studies of learners' inquiry into related variation are concerned with their skills at making fair comparisons in the data to discern causal factors. In these studies, one deductive strategy for making fair comparisons is given primacy in research and instruction – the control-of-variables strategy (CVS). CVS involves making case comparisons that differ on only one independent variable, to see whether that independent variable has an effect on the dependent outcome. Many studies of students' inquiry have demonstrated that CVS can be taught, and that using the strategy can improve students' performance in determining causal factors [17, 18].

However, controlling variables to deduce causes is not the only way scientists make fair comparisons with data. Often such comparisons are made inductively, comparing cases with the same outcome to see if the independent factors that cause that outcome can be determined. This is often the approach in cases when the independent variables cannot be manipulated at will, for instance, in medical diagnosis. Conlin, Hallinen, and Schwartz proposed that this inductive strategy, which they called the general principle strategy (GPS), as an instructional alternative to control of variables. The hypothesis was that GPS would be a superior approach to CVS when learning multivariate relations. Control of variables is not well suited for discovering relations that may only be detectable when multiple quantities are changed at once, since students would have to synthesize the results across many cases in order to find the multivariate pattern. For instance, students using a CVS strategy to investigate the factors that affect how much a spring stretches when a mass is hung from it may systematically vary the length of the spring, the coil diameter of the spring, or the hanging mass one-at-a-time. However, these students may have difficulty finding multivariate effects, such as how the effect of mass on spring stretch depends on the width of the spring, when only changing one variable at a time. GPS may offer a faster route to discovering multivariate relations, by comparing across common outcomes (same amount of spring stretch) and finding common variables or combinations of variables.

To test the hypothesis that GPS would help students detect multivariate relations better than CVS, Conlin, Hallinen, and Schwartz conducted a study in middle-school science. Students in five middle-school science classrooms (n = 100) were randomly assigned into one of two instructional conditions, either general principle (GP) or control-of-variable (CV) instruction. Over 7 class periods, students in each condition were taught their respective inquiry strategy (GPS or CVS), and were taught to apply the strategy in investigating relations for a series of physics topics (projectiles, pendulums, and collisions). In the lessons, students searched for relationships between physical properties through inquiry in hands-on experiments as well as computer-based physics simulations [19].

Strategy use and success were assessed through two post-test items. The "ramp" posttest item presented five cases of balls rolling down a ramp. Each case could differ on one dependent outcome, time to reach the bottom, and three independent variables, ball size, weight, and shape (Figure 1). Students were asked to (1) choose which cases to compare to determine which factor affects the time to roll down the ramp and (2) pick which factors matter, based on the data. The cases chosen were used to classify students' strategies: GPS (selecting cases with the same outcome), CVS (selecting cases that only differed on one independent variable), and neither strategy. Counterintuitively, the shape of the ball was the only factor that affected the time it takes to roll down the ramp.

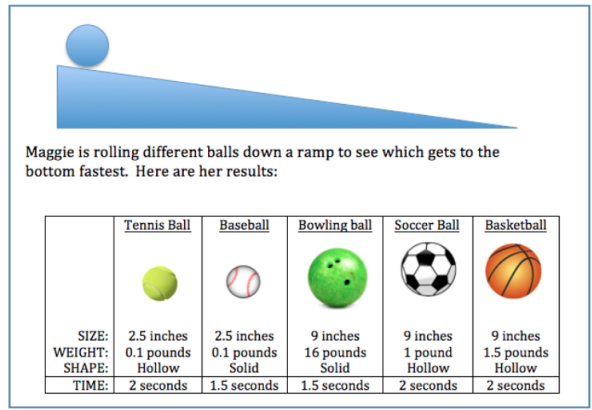

Figure 1. The "ramp" posttest item. In two questions, students were asked to say which balls they would compare and to see if they could figure out what makes a difference in how fast a ball reaches the bottom of the ramp.

The "Balancing Act" posttest allowed students to explore the Balancing Act PhET simulation, in which students could place masses on either side of a see-saw at different distances from the fulcrum. When they were ready, students could choose to enter a challenge mode where they were asked to predict which way different configurations of weights would make the balance beam tip. The most difficult questions required students to know the precise expression for torque (the product of the mass and the distance from the fulcrum) to accurately predict the direction of tipping. In the challenge mode, students answered progressively more difficult questions until they got one wrong or correctly answered 8 in a row. The number of questions correctly answered in the challenge mode was used as a measure of how precisely students' were able to discover the multivariate relation of torque.

The results did not support the research hypothesis that students who used GPS would be better able to detect multivariate relationships in inquiry activities than those using CVS. On the ramp post-test, there was no statistically significant association between condition (CV or GP) and answering correctly, $\chi^2(1, N = 100) = .17, p > .05$. On the balance act posttest, there was no statistically significant difference between the average maximum score for the CV and GP conditions t(63) = 1.08, p = .29.

Although the core prediction of the study was a null result, many subsidiary predictions of the study did support the conclusions that (1) students can learn GPS, and that (2) GPS is a viable alternative strategy for students to discover multivariate relations. For instance, on the Ramp posttest, there was an association between condition (GP or CV) and which strategies students used to make comparisons (GPS, CVS, or Neither) $\chi^2(2, N = 11) = 6.14, p < .01$. A histogram of strategies used in each condition reveals that students were much more likely to use the General Principle Strategy if they were in the GP condition (Figure 2). Likewise, they were much more likely to use the Controlling Variables Strategy if they were in the CV condition. This suggests that GP students did learn to use the general principle strategy, which serves as an existence proof that GPS is teachable.

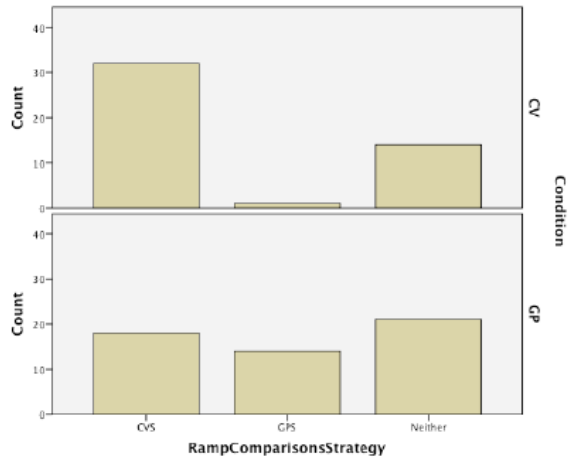

Figure 2. Histograms of comparison strategies used by condition.

Furthermore, there was a significant association between using either inquiry strategy (GPS or CVS) and drawing the correct conclusion the variable that determines how long a ball takes to roll down a ramp $\chi^2 (1, N = 100) = 7.29, p < .01$. Further breaking out Ramp Conclusions by Ramp Strategy shows that either strategy (CVS or GPS) was more effective than using neither strategy for identifying the relevant factor. Although there is no difference in the effectiveness between CVS and GPS strategies, the results show that both are equally successful here.

Table 1. Relative likelihood that each comparison strategy would lead to correctly choosing the relevant factor

| Strategy used on the ramp post-test | % choosing the relevant factor on ramp post-test |
|---|---|
| CVS (N = 50) | 76% |
| GPS (N = 15) | 73% |
| Neither (N = 35) | 49% |

These results also suggest reasons why the overall condition effect on solving the Balance Act Challenge Mode was not significant. There were much fewer students using the GPS strategy on the Ramp Comparisons task (15), compared with CVS (50). It could be that instruction of GPS was not as effective as the instruction of CVS (after all, it was a first attempt to teach the GPS strategy to middle schoolers). Also, the histograms in Figure 1 reveal that while many students in both conditions used CVS, only students in GP condition used the General Principle Strategy. This suggests that the students may have generally been familiar with the strategy of controlling variables, which cut into the uptake of GPS.

Since there was unequal uptake of the strategies by condition, it is informative to see whether those who did learn to apply GPS fared better on the Balance Act Challenge Mode. Figure 2 shows the mean score for students on the Balance Act Challenge Mode (how many scenarios in a row they predicted correctly), broken out by strategy use on the Ramp comparison task. An analysis of variance with the ramp comparison strategy used (GPS, CVS, or neither) crossed with the maximum score on the balance act posttest was significant, $F(2,98) = 6.143, p = .003$. However, a direct comparison of those who used GPS or CVS shows there was no difference in the maximum balance act score t(63) = 1.08, p = .29. This suggests that students who used either strategy on the Ramp Comparison task on average scored significantly higher on Balance Act, compared with those who did not use either strategy, but neither strategy was more effective than the other (Figure 3).

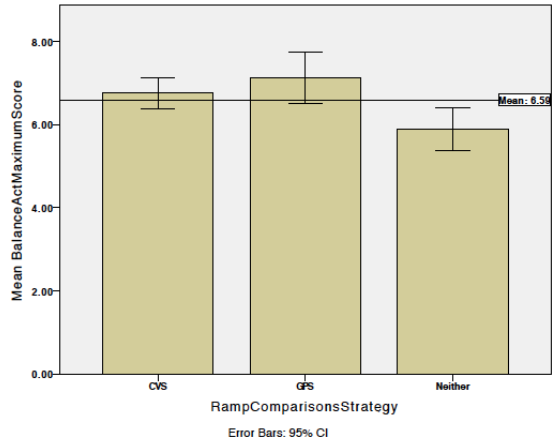

Figure 3. The mean maximum number of challenges answered correctly in Balance Act simulation by strategy used on the "ramp" posttest question. Error bars represent +/- 1 SE.

Although a difference between GP and CV instruction on students' success on the ramp post-test was not found, the results help establish the potential use of a new strategy, the general principle strategy, in teaching science inquiry. Given the prevalence of CVS as an instructional inquiry strategy, it is useful to compare its efficacy to alternatives. Although there was a theoretical reason to believe that GPS would be more successful for detecting multivariate relationships, that was not found to be true here. Yet, GPS was as effective as CVS in science inquiry tasks for detecting new relations. These results established the viability of GPS of a potential alternative to the ubiquitous control-of-variables strategy and could provide an initial step towards finding areas where GPS can be more effective than alternative strategies. They also provided a potential explanation for why the hypothesized benefit of GPS was not found here: because few students took up GPS, it could be that the demonstrable efficacy of the strategies might depend on the effectiveness of the GP instruction. One future direction for revising this study could be to enhance the GP instruction, so that more students take up GPS.

### B. Null results can show how effects fail to generalize to new contexts

Null results can also contribute to new understanding if they show how established effects can fail to generalize to new contexts. In a recent example, Milam, Cohen, Mueller, and Salles [20] showed that stereotype threat did not affect female surgical residents. Stereotype threat occurs when a person in a stereotyped group's feels at risk at conforming to those stereotypes. A well-established version of the stereotype threat effect is that performance can be lowered by reminding the members of the stereotyped group that they are expected to perform worse. Even though a stereotype that women are worse surgeons exists, the standard stereotype threat induction did not lower female surgical residents' performance on mental rotation or working memory tasks. To explain this finding, they propose that female surgical residents are a unique population, who have achieved success in their profession. They propose that this high achieving population may be intrinsically resilient to stereotype threat or may have developed adaptive strategies to mitigate stereotype threat. This study shows that widely established effects may not generalize to a novel population and understanding why they do not may lead to new understandings of a familiar phenomenon.

In our own work, Hallinen and Booth (under review) showed that a mathematical problem-solving effect did not generalize to a new class of problems. Previous research on mathematical problem solving had revealed a verbal advantage for simple algebraic equations [21]. When students solved simple one- and two-step algebra word problems, they were more accurate than when they solved isomorphic problems presented as equations to be solved. In contrast, researchers found a symbolic advantage for more complex algebra problems, such that learners were more accurate when solving equations than equivalent word problems [22]. In these previous experiments, two possible mechanisms were proposed for this shift: (1) the algebraic formalism is more beneficial for more complex problems and (2) students are less likely to use informal solutions when solving complex world problems. Hallinen and Booth (under review) extended this line of inquiry to a new problem type, algebraic proportions. The researchers investigated whether these problems would show a verbal or symbolic advantage.

The study involved students enrolled in Algebra I classrooms (n = 351) throughout the Midwestern US. These students completed a variety of mathematics problems during class time. Six problem types were given to the students: simple and complex linear equations, proportions equations, proportions word problems, and two types of problems that required students to write proportions from word equations ("6 is to 30 as 5 is to x") and from word problems. The primary research question was whether proportion problems would show a verbal or symbolic advantage, which was investigated by comparing accuracy on proportion equation problems and proportion word problems. Secondary research questions included comparison of the relative difficulty of the three types of equation problems: simple algebra, complex algebra, and proportion, and explored students' problem-solving strategies on the proportions word problems.

The primary finding was a null result: there was no significant difference in accuracy between proportions problems presented as equations versus as word problems. Students were generally quite accurate on both of these types of problems (81% correct (SD = 31) for proportions equations and 75% correct (SD = 38) for word problems). Even though the primary result was a null result, this study extends a finding on mathematical problem solving to a new problem type, situating proportion problems into the landscape of types of algebra problems for which this difference has been mapped.

Furthermore, this result helps to build theory regarding the mechanism for symbolic and verbal problem-solving advantages. The secondary analysis showed that accuracy on proportion equation problems was equal to accuracy on simple algebra equation problems and greater than accuracy on complex algebra equation problems. Because proportion equation problems were found to be of equal difficulty to simple equations, the null result for proportion problems shows that problem difficulty alone cannot explain verbal advantages. Another possibility is that the verbal advantage for word problems depends on the availability of informal strategies that can take advantage of the story context of word problems. Investigation of students' written strategies provided support for this explanation by showing widespread adoption of the algorithmic cross-multiplication strategy on proportion problems. Very few students used informal strategies on proportions. The widespread use of this formal algorithmic may have overshadowed informal strategies that would potentially support a verbal advantage for proportion problems. In this study, the null result for proportion problems was itself surprising, but additional analysis helped generate potential explanations that increased our understanding of not only the null result for proportion problems, but also of the mechanisms underlying the original verbal and symbolic advantages seen on algebra problems. Here, situating the null result against the backdrop of existing findings helps bound the generalizability of the existing effects while also providing additional insight into the underlying mechanisms, mapping out the theoretical terrain.

### C. Null results that fail to replicate effects can bring attention to the necessary conditions for replication

Although establishing generalizability is one of the goals when using empirically rigorous methodologies, generalizability is directly supported through replications in different contexts. Null findings in replication studies can help counter the overgeneralization of results and push towards a more nuanced understanding of the mechanisms and necessary conditions needed to reproduce them. This was the case of Kuo, Hallinen, and Conlin [23,24], which set out to replicate a surprising effect of prompting students to draw a diagram during problem solving.

    Free-body diagrams are a ubiquitous instructional support in introductory physics classrooms, intended to help students formalize and keep track of their reasoning on Newton's laws problems. Heckler [25] investigated the effect of prompting students to draw free-body diagrams before solving force problems. Undergraduate introductory physics students at a large midwestern university were given a small set of typical force and motion problems. The students were randomly selected to either receive a version of the questions that included a written prompt to draw force diagram to help solve the problem or to receive no such prompt. As with most problem-solving hints, instructional wisdom would suggest that this hint would help students produce physically correct solutions. Surprisingly, students who were prompted to draw a free-body diagram performed worse on the force problems. By analyzing their written solutions, Heckler found that the prompts led students to engage in formally taught problem-solving methods rather than informal approaches. In alignment with the algebra word problem studies described previously, he proposed that students' may be more successful on force problems when using intuitive, informal solutions rather than formally taught ones.

    Kuo, Hallinen, and Conlin [24] attempted to replicate and extend Heckler's study. The participants were 136 undergraduate students in an introductory algebra-based physics course at a large, private university. In the replication portion of the experiment, students were given two force problems, randomly given a prompt to draw a diagram (n = 66) or no prompt (n = 70) before solving the problem. The key measures were whether students answered the questions correctly and whether they used a formal approach or an informal shortcut.

    The results partially replicated those of the original study: students who were prompted to draw force diagrams used a more formal problem-solving approach, while unprompted students were more likely to use an informal, but more efficient approach. However, unlike Heckler's study, there was no difference in correctness by condition, with accuracy across conditions on problem 1 at 38% and accuracy on problem 2 at 55%. In seeking an explanation for the failure to replicate the effect of prompting on accuracy, we compared the accuracy on problem 1, a problem common to both studies, found in the original and replication studies. The accuracy for the prompted groups in the original and replication study were comparable, but the accuracy for the unprompted, control group was about 20% higher in Heckler's original study. We proposed that one explanation could be due to the unusually fast pace of the physics course in the replication study, which spent only 2 50-minute lectures covering force problems. Additionally, the replication study uncovered a possible mechanism for how prompting could lower accuracy. In the replication study, it was found that solution type used affected accuracy, where the formal Newton's 2nd law approach was less accurate than informal alternatives. Most incorrect, formal solutions were incomplete; students knew how to start the Newton's 2nd law approach, but could not complete it to solve for the relevant force. Although the overall effect was not replicated, these results suggested that prompting could lower problem solving success by suggesting formal solution approaches that students did not know how to complete.

    Kuo, Hallinen, and Conlin [24] set out to replicate Heckler's surprising result that physics students who were prompted to draw free body diagrams performed worse on Newton's laws problems [23]. While the effect of prompting on correctness was not replicated, comparison to the results in the original study suggested a possibility for the discrepancy: the difference in accuracy may depend on how much time students are given to learn force problems before moving on to the next topic. Effects on accuracy may not exist if students in the class have not achieved a certain level of problem-solving skill. Here, the failure to directly replicate a result bounded the generalizability of that result and suggested a possible necessary condition for its replication.

## IV. RESEARCH TO GENERATE NEW UNDERSTANDING: SUGGESTIONS FOR EXPERIMENTAL DESIGN AND EVALUATION

Our primary goal in this paper is to challenge common criteria for evaluating the quality of research in the culture of research fields. The quality of a quantitative study is often judged according to whether it shows novel effects and whether it establishes those effects with statistically significant results - that is, whether the results were unlikely to have been sampled from a null hypothesis distribution. Here, we propose an alternative measure of quality: whether a quantitative study has the potential to generate new understanding. Using this criteria, even null results can be seen as important research results. We have presented three ways that null results can contribute to new understanding: by providing existence proofs of the viability of new approaches, by showing how previous results can fail to generalize to new contexts, and by uncovering the necessary conditions for replication.

To be clear, we are not suggesting that all null results should be published. To flood journals with underpowered designs and idiosyncratic failures to replicate a finding would be impractical at best and has the potential to drown out new understandings with noise. What we propose is to evaluate research that produces null results according to its potential for generating new understanding. But how can this be accomplished practically? One approach that is gaining traction in psychology is the review of pre-registered studies [26]. In a pre-registered study, researchers specify the research questions, experimental design, and planned analysis beforehand. Some journals are reviewing these pre-registered studies as registered reports, with accepted studies being given the promise of publication independent of the findings. This approach to reviewing studies based on their design, the importance of the research question, and potential contribution to the literature, without knowing whether the results will show statistically significant patterns, aims to reduce the publication bias against null results.

We also argue that replication studies, in addition to attempting to replicate an existing finding, should also seek to uncover new explanatory mechanisms or connections to existing research areas. In the cases of published null results in our previous work, analysis of additional measures provided an opportunity to develop and support new explanations for the data. We suggest that the most useful replication studies can enhance their contribution by designing for enhanced understanding. In these cases, the understanding of a null result can be enhanced when situated in a network of other results. Therefore, we suggest that researchers employ a suite of process measures in their experimental designs.

Large gains on summative assessments can be a good sign of a successful instructional intervention, but without corroborating against process data and other assessments it can be unclear why the intervention was successful. In the event of a possible null result, these corroborating measures can be vital in understanding two things: (1) if there was a result on some outcome of interest even if the overall research question led to a null result, and (2) what may have contributed to the fact that the results were null, either as a manipulation check or by analyzing why the treatment may not have had the intended effect on students' engagement or activities throughout the intervention. Even if the study findings are not a null result, having additional measures can only add to the richness of the research conclusions and provide more insight into the mechanisms by which the finding may have occurred. Careful articulation of research designs can help make clear what measures will be needed to answer particular process questions in the research study. Although much analysis can be done post-hoc, the most important consideration in designing the study is which measures to include.

Beyond study design, we suggest that presentation of null results should be particularly attentive to framing that highlights the contribution of the null result. This framing includes (1) situating the importance of the null result in challenging or extending existing theories and empirical results and (2) presenting process measures that provide insight into the mechanisms and boundaries underlying the effect under investigation. In addition, we suggest that researchers share as much as possible about their materials and results, to facilitate discussion among the research field as well as future related studies. PRPER publishes supplemental materials that can contain more complete versions of study materials. In addition, the growing practice of open science repositories can allow researchers to upload detailed information about their study, including raw data and analysis methods.

## V. DISCUSSION

There are at least two reasons to publish null results in an educational science such as PER. One reason is that science is an inductive process. In order to develop our understanding, we need a complete, unbiased sample of research results across which we can find patterns and reliable effects. Another reason is that science is the process of searching for mechanism. Null results, when situated in the existing literature and in a network of additional process measures, can be a generative source for establishing new mechanisms in teaching and learning. Yet, as with many research cultures, there continues to be a publication bias against null results. As we have discussed, this publication bias runs the risk of producing a bias in our understanding of educational phenomena.

In the classrooms we study, we are increasingly calling on our students to embrace mistakes as part of the learning process, even part of the *scientific* process. By devaluing null results, we risk not practicing what we preach. If we view null results as "mistakes" in that they reflect our inability to make valid research predictions or support those predictions empirically, then we should not be afraid to embrace our "mistakes" and learn from them. The common practice of keeping null results in our private file drawers reflects a lack of openness to learn from feedback. To overcome this problem, authors, reviewers, and journal editors alike will need to shift their focus from the statistical significance of results to their significance for enhancing our understanding.